\documentclass[conference]{IEEEtran}
\IEEEoverridecommandlockouts
\usepackage{cite}
\usepackage{amsmath,amssymb,amsfonts}
\usepackage{algorithmic}
\usepackage{graphicx}
\usepackage{textcomp}
\usepackage{xcolor}
\usepackage{soul}
\def\BibTeX{{\rm B\kern-.05em{\sc i\kern-.025em b}\kern-.08em
    T\kern-.1667em\lower.7ex\hbox{E}\kern-.125emX}}
\begin{document}

\title{Smarter Password Guessing Techniques Leveraging Contextual Information and OSINT}

\author{\IEEEauthorblockN{Aikaterini Kanta\IEEEauthorrefmark{1}\IEEEauthorrefmark{2}, Iwen Coisel\IEEEauthorrefmark{2}, Mark Scanlon\IEEEauthorrefmark{1}}
\IEEEauthorblockA{\textit{\IEEEauthorrefmark{1}Forensics and Security Research Group, School of Computer Science, University College Dublin, Ireland}}
\textit{\IEEEauthorrefmark{2}European Commission, Joint Research Centre (DG JRC) - Via Enrico Fermi 2749, 21027 Ispra (VA), Italy}\\

Email: aikaterini.kanta@ucdconnect.ie, iwen.coisel@ec.europa.eu, mark.scanlon@ucd.ie}


\maketitle

\begin{abstract}
In recent decades, criminals have increasingly used the web to research, assist and perpetrate criminal behaviour. One of the most important ways in which law enforcement can battle this growing trend is through accessing pertinent information about suspects in a timely manner. A significant hindrance to this is the difficulty of accessing any system a suspect uses that requires authentication via password. Password guessing techniques generally consider common user behaviour while generating their passwords, as well as the password policy in place. Such techniques can offer a modest success rate considering a large/average population. However, they tend to fail when focusing on a single target -- especially when the latter is an educated user taking precautions as a savvy criminal would be expected to do. Open Source Intelligence is being increasingly leveraged by Law Enforcement in order to gain useful information about a suspect, but very little is currently being done to integrate this knowledge in an automated way within password cracking. The purpose of this research is to delve into the techniques that enable the gathering of the necessary \textit{context} about a suspect and find ways to leverage this information within password guessing techniques.
\end{abstract}


\section{Introduction}
Nowadays, criminal activity is increasingly conducted in cyberspace. Criminals take advantage of the easy access to information and the global access to victims. This leads to a scale of crime that cannot be easily achieved in the physical world. As a result, it is becoming increasingly urgent for law enforcement to be able to act swiftly in a digital forensic investigation especially in the cases where ongoing or future criminal acts must be prevented. Very often, password protected accounts or encrypted devices act as a barrier for police personnel to conduct their lawful investigations~\cite{SAYAKKARA201943}.

Passwords have been the go-to method of user authentication for decades -- a fact that does not look like it is about to change. The difference in the last few years is the fast increase of online login systems with password policies that require passwords of different patterns, lengths and makeup~\cite{ur2012does}. This leads to users having the tendency to either reuse the same password across different systems or to create passwords that are easy to remember (i.e., weaker) to keep up with the different password policies~\cite{komanduri2011passwords, lord_2018}. A significant portion of passwords are therefore based on dictionary words~\cite{yan2004password} or contextual information related to the user~\cite{wang2016targeted}. Taking this into account, this work is exploring ways this information about a suspect as an individual or a community can be leveraged in order to better facilitate the recovery process.

\section{Related Work}
\subsection{Password Metrics}
There have been studies that analyse the composition of passwords with focus on how people choose them, where password re-use and reliance on dictionary words can be observed~\cite{yan2004password, lord_2018, pearman2017let}. In addition, studies have considered the demographics of participants and if they play an important role in the selection process~\cite{bonneau2012science, alsabah2018your}. Finally, it is observed that users tend to use personal information when they create a password, as it is more easily memorable to them~\cite{wang2016targeted,li2016study}.

\subsection{Open Source Intelligence} 
\label{osint}
Steele \cite{steele2007open} defines Open Source Intelligence (OSINT) as information that is publicly available and can be used to answer a specific question. To this end, there are many OSINT tools available to the community that can aid in finding, extracting and sorting though this information\footnote{https://osintframework.com/}.

\subsection{Password Guessing Related Tools}
\label{pgt}
Traditional password guessing techniques include brute force, dictionary attacks and rainbow table approaches. Lately, newer, smarter methods have been proposed with higher recovery numbers, e.g., password candidate generation tools based on Markov Chains~\cite{durmuth2015omen}, probabilistic context-free grammars~\cite{weir2009password} and variations of thereof~\cite{li2016study}, and combinator attack tools, e.g., PRINCE\footnote{https://github.com/hashcat/princeprocessor}.

\section{Methodology}
Digital forensic investigators frequently find themselves working on a case where a password connected to a crime needs to be retrieved. Many times, it is not possible to retrieve the password of a suspect in a timely manner. This is where tying the suspect to their associated \textit{contextual} information can prove fruitful. Context in this instance refers to contextual information about a suspect that can be harnessed for the sake of making better ``educated" guesses about their password.

\begin{figure}[htbp]
\centering
\includegraphics[width=0.25\textwidth]{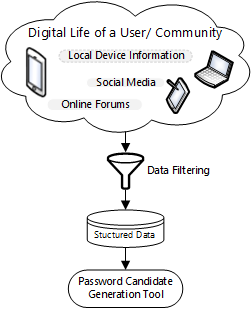}
\caption{Password Generation Process}
\label{fig:model}
\end{figure} 

This information can stem from their online/offline life and can be the product of traditional means of investigation, e.g., forensic investigation of the suspect's residence and belongings or the digital investigation of the suspect (i.e., both their local devices and online presence). The latter type of investigation can yield results including the suspect's interests (e.g., sports, music, etc.), browsing history, other passwords, online interactions, family and pet names, etc. This is where OSINT can play a key role in an investigation. Already Social Media Intelligence (SOCMINT) is used by law enforcement in order to gain pertinent, case-progressing information~\cite{van2009introduction}. It can be applied to online groups of users to detect patterns in social behaviour as well as to individual suspects~\cite{ivan2015social}. OSINT tools can extract information regarding the online presence of a user or group of users from networks of acquaintances on social media. The prevalence of users increasingly living their lives online can also result in sourcing their email addresses, usernames, phone numbers, and exercise or sleeping patterns. 

The information gathered can then be leveraged, filtered and translated to meaningful contextual data about the suspect. Information retrieval and machine learning techniques should be beneficial at this stage, based on the volume of available data and whether an individualised or community-based approach is chosen. For the individual-based approach, information gathered online and offline will be compiled. An analysis on the raw data and a classification into categories is the next step, in order to extract meaningful keywords that will represent user/community interests. In turn, these keywords will help law enforcement officers assemble a more useful list of password candidates to enrich and complement existing password guessing tools. A flowchart of this process can be seen in Figure~\ref{fig:model}.

\section{Conclusion and Future Work}

To validate the proposed hypothesis, i.e., prove that context does play a role in the selection of passwords, our future work will firstly focus on a community-based approach. The reason for this, is the current lack of available data for individuals and the sensitivity of this data~\cite{fleurbaaij2017p}. As one example, a community of users on an anime forum would be expected to have a higher percentage of passwords related to anime than a community about cooking. Current work is focused on the contextual analysis of a large corpus of passwords stemming from online leaks. Future work will include the refinement of this dataset and the expansion of the scope of this analysis to more contextual information. In addition, the manner with which OSINT can be exploited and processed towards producing meaningful data will be explored. This will subsequently be a starting point for creating a bespoke, personalised dictionary list to feed into password cracking tools. 

\bibliographystyle{IEEEtran}
\bibliography{bibfile}
\end{document}